\documentclass[]{spie}  

\usepackage{amsmath,amsfonts,amssymb}
\usepackage[numbers,sort&compress]{natbib}
\usepackage{graphicx}
\usepackage{hyperref}
\usepackage{etoolbox}
\apptocmd{\thebibliography}{\setlength{\itemsep}{-2pt}}{}{}
\usepackage[justification=centering]{caption}

\title{Fader Networks for domain adaptation on fMRI: ABIDE-II study}

\author[a]{Marina Pominova*}
\author[a]{Ekaterina Kondrateva*}
\author[a]{Maxim Sharaev}
\author[a]{Alexander Bernstein}
\author[a]{Evgeny Burnaev}
\affil[a]{Skolkovo University of Science and Technology}

\begin{document} 
\maketitle

\begin{abstract}
ABIDE is the largest open-source autism spectrum disorder database with both fMRI data and full phenotype description. These data were extensively studied based on functional connectivity analysis as well as with deep learning on raw data, with top models accuracy close to 75\% for separate scanning sites. Yet there is still a problem of models transferability between different scanning sites within ABIDE. In the current paper, we for the first time perform domain adaptation for brain pathology classification problem on raw neuroimaging data. We use 3D convolutional autoencoders to build the domain irrelevant latent space image representation and demonstrate this method to outperform existing approaches on ABIDE data.
\end{abstract}


\section{INTRODUCTION}
\label{sec:intro}  

\subsection{Autism spectrum disorder classification}


Autism spectrum disorder (ASD) is wide diagnosis capturing different developmental disorders \cite{postema2019altered}. This disorder is being widely studied now by separate laboratories and consortiums resulting in several publicly available ASD neuroimaging datasets. 

Modern machine and deep learning models are one of the most powerful techniques to extract information from complex and high dimensional fMRI imaging data. However, due to the difficulty of collecting medical data, the typical size of such datasets usually varies within a few tens or hundreds of observations, which is too small for most statistical instruments and, most importantly, proper training of machine learning and deep learning models. The results of neuroimaging studies of ASD have often been inconsistent due to small sample sizes in relation to subtle effects \cite{postema2019altered}. 

One of the possible solutions is the meta-analysis of several datasets collected for a single research task. Unfortunately, in this case, another problem arises: due to differences across studies (such as differences in methodology, equipment, scanner settings, processing pipelines, etc.), the obtained scan data vary in imaging properties, which significantly affects the generalization ability of the resulting models and might nullify the meta-analysis benefits. Moreover, the applicability of the derived models to the new test data from other sources is also usually hurt due to inter-site variability effects.

Autism Brain Imaging Data Exchange (ABIDE) initiative is the largest open-source data collection, which has two releases: ABIDE-I (2014) and ABIDE-II (2017). To date, ABIDE-II has collected over 1000 additional studies, while the total number of observations in the two releases has exceeded 2000 subjects from 29 different sites. The difference between sites is not only in processing pipeline or patient cohort but also in scanning equipment ($1.5$T or $3$T MR machines). Analysis of fMRI data collected in the scope of ABIDE initiative may help to shed light on the causes  and mechanisms of ASD, as well as to identify objective biomarkers of this disorder which could facilitate diagnostics, therapy and outcome predictions [2]. However, to get the most out of the analysis of this large and diverse dataset, the difficulties associated with inter-site data variation and resulting low robustness and generalization ability of analytical models have to be overcome.

\subsection{ABIDE: previous studies}

To the moment, there exists a large scope of research focused on the study of the first phase ABIDE - ABIDE-1 dataset. However, in most cases, the authors analyzed small subsets from separate sites, separately or as part of a joint sample, but without trying to exclude the cross-site variability effects. Several ASD detection models trained on small samples achieved classification accuracies up to 75\% \cite{yahata2016small}. At the same time, the majority of studies conducted on the whole multi-site dataset showed significantly lower accuracy around 65\% \cite{abraham2017deriving}  \cite{el2019hybrid}.

The studies also differ methodologically: more frequently they use functional connectivity analysis \cite{kazeminejad2019topological}, while others apply deep learning approaches directly to raw or preprocessed data, see \cite{price2014multiple} \cite{el2019hybrid}.  While working with vectorized features from fMRI functional connectivity matrix is not new, some authors explore graph-theoretic and topological features from connectivity graphs and report comparable or better performance (around 65-80\% accuracy depending on patient cohort) than traditional approaches \cite{kazeminejad2019topological}.


The first application of networks with LSTM is demonstrated in \cite{cheng2017classification} to process raw fMRI time-series data from a multi-site heterogeneous  ABIDE-I  dataset with an accuracy of 68.5\%  on cross-validation. It is important that the authors also made one more step towards revealing autism biomarkers from fMRI data, trying to interpret the best model and identify brain regions important for distinguishing subjects with ASD from typical controls

With the ability to automatically learn the task-specific features from unstructured image or sequence data, deep learning techniques also have the potential to be applied to raw full-size fMRI data. Such architectures already demonstrated a possibility to detect various psychoneurological conditions, for example, epilepsy, schizophrenia, and major depressive disorder from full-size fMRI without any manual feature extraction [Dakka, Pominova]. Several modifications of the neural network architecture are made to capture the one-dimensional temporal and three-dimensional spatial properties of 4D fMRI data, for example, in \cite{el2019hybrid}. Here, in the experiment with several sites, the accuracy is significantly (64\% versus 78\% for the best model) lower than in the experiment with a single site, which is consistent with observations in other works. An interesting combination of biologically plausible feature extraction and deep learning approach is demonstrated in \cite{khosla2019ensemble}. Here the input to the 3D CNN is formed by concatenating channel-wise several three-dimensional voxel-level maps of functional connectivity, calculated from raw fMRI time-series. Furthermore, the authors applied the saliency map approach to their 3D CNN model to visualize discriminative features and assumed that unsupervised learning on these maps can provide insights into the clinical origins of disease \cite{khosla2019ensemble}.

The small number of exceptions in which there have been attempts to explicitly account for site-related differences and data heterogeneity include, for example, work \cite{postema2019altered}, in which the authors fitted linear mixed-effects models for each calculated feature with respect to the data from 54 data sets (data sets were coded as factor variables). Federated learning and domain adaptation methods were proposed in \cite{li2020multi} to train the models locally on separate sites data and to account for differences of fMRI signal distributions from different sites.  Specifically, a mixture of experts (MoE, adaptation near the output layer), and adversarial domain alignment (adaptation on the data knowledge representation level), significantly increased the model accuracy, see \cite{li2020multi} for details. It is reasonable to assume that reliable biomarkers could be detected from a reliable (robust across sites) model. Thus, federated learning might be helpful to detect replicable and robust biomarkers across different sites.


In the present study, we focus on the problem of eliminating site-related differences from fMRI data and propose an approach to train robust and transferable between sites neural network models on a multi-site dataset. We conduct experiments on the data collected in the scope of the ABIDE initiative. Since the first phase of ABIDE (ABIDE-I) is already a well-studied dataset, here we focus on the recently released second phase (ABIDE-II). 
First, for the four data sites with largest contribution in the ABIDE-II dataset, we provide ASD recognition baselines obtained with two widely used approaches — conventional functional connectivity matrices analysis and recently proposed full-size MRI series analysis with a conv-temporal neural network \cite{dakka2017learning}\cite{pominova2018voxelwise}. We present classification results on the merged dataset of all 4 sites, as well as on each site data separately.  
Second, we propose a novel approach based on 3D convolutional autoencoders, fader networks and domain-adaptation techniques to remove site-related variability and reduce the heterogeneity of the data from different sites. To validate if the proposed method allows enhancing the generalization ability of ASD classification models, we compare it with baseline conventional methods for connectivity matrices and full-size fMRI on the same data sample.




\section{Methods}

\subsection{Data and preprocessing}

We perform the study on the publicly available ABIDE-II dataset\footnote{http://fcon\_1000.projects.nitrc.org/indi/abide/abide\_II.html} which contains resting-state fMRI sequences of patients with ASD  and healthy controls \cite{di2017enhancing}.  We use the whole data sample from 4 collection sites, totaling 352 subjects. The distribution of ASD subjects per-site represented in the Table \ref{tabl2}.

\begin{table}[t!]
\caption{ABIDEII distribution of ASD subjects per-site }\label{tabl2}
\centering
\begin{tabular}{|l|l|l|l|l|l|} \hline 
\begin{tabular}[c]{@{}l@{}}\textbf{Description}\end{tabular}
 & \begin{tabular}[c]{@{}l@{}}  \textbf{NYU} \\ (NYU\_1, NYU\_2)\end{tabular} & \begin{tabular}[c]{@{}l@{}}\textbf{OHSU} \\ (OHSU\_1)\end{tabular} &  \begin{tabular}[c]{@{}l@{}}\textbf{KKI} \\ (KKI\_1)\end{tabular}  & \begin{tabular}[c]{@{}l@{}} \textbf{GU}\\ (GU\_1)\end{tabular} & \textbf{All sites} \\
 \hline
Total Subject & 105 & 93 & 79 & 75 & 352 \\ \hline
ASD Subject & 30 & 56 & 62 & 36 & 184 \\ \hline
Mean age (STD) & 10.01 (2.4) & 9.60 (4.61) & 10.20 (4.02) & 10.64 (1.72) & 10.15 (2.98) \\ \hline
\end{tabular}
\end{table}

The sequences of images were preprocessed according to Functional Connectomes Project / INDI \cite{craddock2013neuro} protocols which include slice timing correction,  motion realignment,  intensity normalization, and rigid registration to a template.  We use the denoised dataset, which is as for now the only available option with samples from all collecting sites.

For Functional Connectivity (FC) analysis we applied the standard \texttt{Nilearn} pipeline\footnote{https://nilearn.github.io/modules/generated/nilearn.input\_data.NiftiLabelsMasker.html} with band-pass filtering \texttt{lowpass} $= 0.1$ and \texttt{highpass} $= 0.01$ Hz,  signal standardization, and de-trending.  For brain parcellation, we use an Automated Anatomical Labeling (AAL) atlas consisting of 116 regions.  For each region, the corresponding time series is taken as the first component of Singular Value Decomposition with \texttt{high variance confounds} function of all the region voxels. Next, the functional connectivity matrix is computed as a $116 \times 116$ matrix of pairwise correlations of the obtained time series for all regions.

\subsection{Fader network for elimination of site-related information}

Various domain adaptation techniques are proven candidates to perform training of domain-invariant models with enhanced generalization ability. One of the most widely used methods is a DANN approach proposed by \cite{ganin2015unsupervised}. This allows adapting a pretrained model to a new domain by fine-tuning it with an additional network, the domain discriminator. The model is trained for a target task only on a sample from “source domain”. At the same time, the discriminator takes features extracted by the model from both "source" and "target" domain data and tries to distinguish which domain these features come from. Usually, these features are taken from the last hidden layer of the model. The main model is trained with gradient descent in a standard way. However, when the discriminator is optimized, the gradients backwarded through the weights of the main model are reversed. As a result, the model learns to extract the features that confuse the discriminator most and complicate distinguishing between domains.

However, a potential drawback of this approach in the case of complex and high-dimensional functional MRI data is the difficulty of stabilizing adversarial training. Our preliminary experiments revealed that when the domain adaptation approach is applied during model training for the target task, the main classification network converges too fast and begins to overfit. Thus, the domain discriminator does not have enough time to properly guide the learning of domain-invariant features.

For this reason, we opted for applying a similar technique and train a fader network \cite{lample2017fader} - an autoencoder-like model for preliminary extraction of site-independent features, which can be further used for training an ASD-recognizing classifier.

Fader networks were suggested as a method for training a conditional autoencoder for generation images with predefined values of some attributes, such as age, hair color, and so on. The trained model is able to produce different versions of input image conditioned on the chosen attribute values. This is achieved by disentangling the salient information of the image and the values of attributes directly in the latent space. The decoder takes desirable attribute values as an additional input along with the latent code and generates an image with the corresponding characteristics. At the same time, the latent representation itself remains independent and does not contain any attribute-related information. 

The elimination of the information on attributes from the latent space of the autoencoder is achieved in a manner similar to the domain adaptation approaches. During training the encoder-decoder model, an additional discriminator network tries to predict attribute values of each training image from its latent representations. The latent discriminator is optimized adversarially with the encoder, forcing it to filter out the information related to the attributes. 

Thus, the entire model consists of three main components:
--- the encoder $E$, a convolutional network that maps the input image $x$ to its latent representation $z(x)$,
--- the decoder $D$, a deconvolutional network that takes as input latent representation $z$ of the image and a vector of attribute values y, and generates a new version of the image with visual characteristics corresponding to attributes - $x'(z, y)$,
--- the latent discriminator $L$, which tries to identify true attribute values of the input image x based on its latent representation $z$. The discriminator is optimized adversarially with the encoder-decoder architecture, and to make it unable to predict the right attributes, the encoder has to learn attribute-invariant representations.

In this study, we apply the Fader networks approach to obtain an encoder capable of extracting site-independent latent representations from full-size fMRI data. We train the encoder-decoder model on 3D brain images sampled from fMRI time series at random time steps and consider site labels as a single attribute variable that the discriminator tries to predict from latent representations.



\begin{figure}[ht]
\begin{center}
\begin{tabular}{c} 
\includegraphics[height=7.8cm]{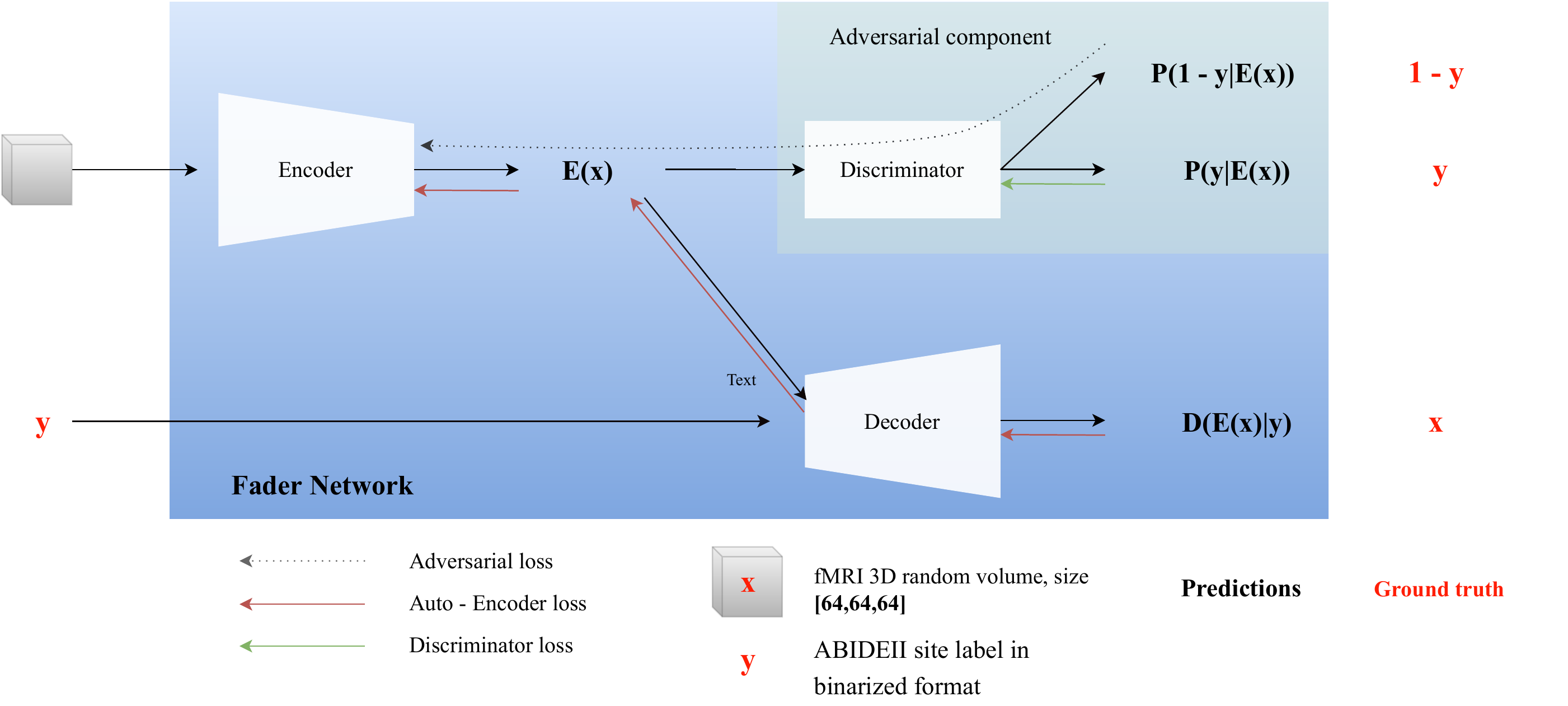}
\end{tabular}
\end{center}
\caption[example] 
{ \label{fig:main} 
 Fader Network architecture used. An fMRI image - $x$ and ABIDE-II site id (one-hot-encoded) - $y$ are given as input. The encoder maps $x$ to it's latent representation; the discriminator is trained to predict $y$ given $z$ whereas the encoder is trained to make it impossible for the discriminator to predict $y$ given vector of latent features only. The decoder should reconstruct $x$ given the latent representation and $y$.}
\end{figure} 

\subsection{Classification of latent representation sequences}

After the Fader Network has converged, we use the encoder to transform the fMRI sequence into a sequence of one-dimensional vectors of latent features. These features are not trained for any specific task and are expected to contain maximum information from the original fMRI series in a compressed form, except for the site-related information, which is excluded by the training procedure. They can be further used to train any predictive model, for example, a classification recurrent network for ASD detection.

It may also be noted that combining the encoder that extracts feature descriptions from brain images in the fMRI sequence with a recurrent classification network that processes the temporal information is architecturally equivalent to the ConvTempNet, a proven architecture for full-size fMRI analysis proposed by [Dakka, Pominova]. This architecture consists of convolutional and recurrent components aimed to take into account both spatial and temporal patterns in the data. During the training process, the convolutional part takes as input 3D brain images at each time step of the fMRI sequence and transforms them into the one-dimensional feature vectors. Thus, the whole fMRI time series is compressed into a sequence of 1D vectors, which is then fed into the recurrent part of the network to extract temporal information.  

Therefore, it is natural to compare the described model with our proposed fader network-based architecture. We also consider a comparison with a model consisting of the pretrained encoder of the plain autoencoder and recurrent classification network to directly assess the effect of site information elimination.

\subsection{Experiment details}

To estimate the influence of site-related information on the performance of ASD detection from full-size fMRI data, we examine the following three architectures:

--- an original ConvTempNet, consisting of the convolutional and recurrent parts,

--- a pretrained encoder of the plain autoencoder connected with a recurrent neural network,

--- our proposed approach with a pretrained site-independent fader network encoder connected with a recurrent neural network.

We also consider two degrees of elimination of the information associated with the data site - the slight adjusting of latent vectors to reduce the share of site-related variability and almost complete its removal of it. 
We denote the former option as FaderNetwork-slight - FaderNetwok with slight site-related information elimination, and the latter as FaderNetwork-strong - FaderNetwok with strong site-related information elimination.

For a uniform and fair comparison, we use an identical convolutional architecture for the convolutional part of ConvTempNet and encoders of the conventional autoencoder and the fader network. Specifically, for the fader net, we adapted a 3D-version of the architecture proposed in \cite{lample2017fader}. The encoder part consists of 6 blocks of convolutional layers with kernel size $4 \times 4 \times 4$, a stride of $2$, and a padding of $1$. The number of channels starts from $32$ and multiplies by $2$ with each next block, while the spatial resolution of the activations reduces twice along each axis.  We also apply leaky-ReLU with a slope of $0.2$ in the encoder, as well as batch normalization after all convolutional layers except for the first one. Each time point of an fMRI sequence is represented by a three-dimensional image of size $64 \times 64 \times 64$.  After going through the encoder, it is transformed into a one-dimensional latent representation with $1024$ feature maps. The decoder part architecture accurately mirrors that of the encoder. The only difference is that here we apply 3D transposed convolutions with the same parameters, and thus each of 6 blocks of layers doubles the spatial resolution of activation maps back to $[64, 64, 64]$. Also, convolutional layers are followed by ReLU activation and batch normalization. The plain autoencoder model has the same architecture and the convolutional part of ConvTempNet is analogous to the encoder.

For the latent discriminator, we construct a small classification network with 3 fully-connected layers of $1024$, $256$, and $64$ units respectively, ReLU activations, batch normalization, and dropout of $0.3$ after each layer. In our experiments, both dropout and batch normalization turned out to greatly influence the success of an adversarial training procedure.

Similarly, we also fix a single recurrent architecture for the temporal part of ConvTempNet and for processing sequences of one-dimensional feature vectors, extracted by autoencoder's and fader network's encoders. Our recurrent network consists of $2$ layers of GRU units with $512$ memory units each. After they processed an input sequence of one-dimensional feature vectors extracted from the fMRI time series, the hidden states of the second GRU layer memory cells are  averaged over time and used to predict the final probability of ASD presence. 



Training of all three models is performed on the same training set and with a fixed validation strategy. We assess the classification performance of each model with the ROC AUC metric computed on $10$-fold cross-validation. Cross-validation splits for all models were identical. We also perform model evaluation on the leave-one-site-out validation. In this case,  we alternately select one site for testing, train the model on the data from all remaining sites and then assess its performance on the data from the selected site that it has not seen before. It allows us to estimate model robustness to the site-related variations in a training set and transferability to the data from new sources. 

\subsection{Classification of functional connectivity matrices}

We also performed a comparison of the proposed models for full-size fMRI data with the classical approach to fMRI analysis based on functional connectivity matrices. 

Feature descriptions for each object were obtained by flattening the upper-triangular part of the symmetric functional correlation matrix into a vector of $6670$ connectivity features. Since their number exceeds the number of objects, feature selection is an important step before training the model. We select the $200$ most important variables according to their weights in the logistic regression model on the training set. Then, these features are used to train the final classification mode on the same training. As a final classifier, we used the SVM algorithm with hyper-parameters optimized with grid search. 

\section{Results}




\begin{table}[t!]
\caption{Results of using different training strategies on data from different ABIDE sites, on 10-fold cross-validation and leave-one-site-out validation, ROC AUC(Std)}\label{tab1}
\centering
\begin{tabular}{|l|llll|l|l|}
\hline
\textbf{Model}                                                                                                & \textbf{GU}                                                                  & \textbf{KKI}                                             & \textbf{NYU}                                             & \textbf{OHSU}                                            & \textbf{All sites}                                                & \textbf{LOSO}                                                    \\ \hline
\text{FC, SVC (kernel = rbf)}                                                                               & \begin{tabular}[c]{@{}l@{}}0.737 \\ (0.120)\end{tabular}                     & \begin{tabular}[c]{@{}l@{}}0.687\\ (0.152)\end{tabular}  & \begin{tabular}[c]{@{}l@{}}0.656 \\ (0.120)\end{tabular} & \begin{tabular}[c]{@{}l@{}}0.539 \\ (0.119)\end{tabular} & \begin{tabular}[c]{@{}l@{}}0.635 \\ (0.051)\end{tabular}          & \begin{tabular}[c]{@{}l@{}}0.579 \\ (0.064)\end{tabular}         \\ \hline
\text{\begin{tabular}[c]{@{}l@{}}ConvGRU \\ on full-size fMRI\end{tabular}}                                 & \begin{tabular}[c]{@{}l@{}}0.543 \\ (0.098)\end{tabular}                     & \begin{tabular}[c]{@{}l@{}}0.586 \\ (0.214)\end{tabular} & \begin{tabular}[c]{@{}l@{}}0.771\\ (0.063)\end{tabular}  & \begin{tabular}[c]{@{}l@{}}0.664 \\ (0.072)\end{tabular} & \begin{tabular}[c]{@{}l@{}}0.665 \\ (0.049)\end{tabular}          & \begin{tabular}[c]{@{}l@{}}0.541 \\ (0.077)\end{tabular}         \\ \hline
\text{\begin{tabular}[c]{@{}l@{}}GRU on latent vectors \\ (Autoencoder)\end{tabular}}                       & \begin{tabular}[c]{@{}l@{}}0.624 \\ (0.101)\end{tabular}                     & \begin{tabular}[c]{@{}l@{}}0.519 \\ (0.225)\end{tabular} & \begin{tabular}[c]{@{}l@{}}0.760 \\ (0.059)\end{tabular} & \begin{tabular}[c]{@{}l@{}}0.646 \\ (0.103)\end{tabular} & \begin{tabular}[c]{@{}l@{}}0.667 \\ (0.075)\end{tabular}          & \begin{tabular}[c]{@{}l@{}}0.534 \\ (0.082)\end{tabular}         \\ \hline
\text{\begin{tabular}[c]{@{}l@{}}GRU on site-invariant \\ latent vectors (FaderNetwok-slight)\end{tabular}} & \begin{tabular}[c]{@{}l@{}}0.611 \\ (0.089)\end{tabular}                     & \begin{tabular}[c]{@{}l@{}}0.579 \\ (0.201)\end{tabular} & \begin{tabular}[c]{@{}l@{}}0.763 \\ (0.065)\end{tabular} & \begin{tabular}[c]{@{}l@{}}0.678 \\ (0.091)\end{tabular} & \textbf{\begin{tabular}[c]{@{}l@{}}0.731\\ (0.078)\end{tabular}}  & \textbf{\begin{tabular}[c]{@{}l@{}}0.652\\ (0.080)\end{tabular}} \\ \hline
\text{\begin{tabular}[c]{@{}l@{}}GRU on site-invariant \\ latent vectors (FaderNetwok-strong)\end{tabular}} & \begin{tabular}[c]{@{}l@{}}0.575\\ (0.093)\end{tabular} & \begin{tabular}[c]{@{}l@{}}0.535\\ (0.170)\end{tabular}  & \begin{tabular}[c]{@{}l@{}}0.729\\ (0.049)\end{tabular}  & \begin{tabular}[c]{@{}l@{}}0.637\\ (0.161)\end{tabular}  & \textbf{\begin{tabular}[c]{@{}l@{}}0.692 \\ (0.094)\end{tabular}} & \textbf{\begin{tabular}[c]{@{}l@{}}0.597\\ (0.091)\end{tabular}} \\ \hline
\end{tabular}
\end{table}

In this section, we demonstrate the results of our experiments. The Table \ref{tab1} shows ROC AUC scores of the three models for full-size data and the baseline model for functional connectivity features. For each model, we present its classification performance on small datasets from individual sites, on the joint dataset from all 4 sites, and on leave-one-site-out validation.

When analyzing the results, we primarily focus on a) the impact of the information associated with the data sites on the model generalization, b) the effect of removing the site-related information on the generalization ability and transferability of the model.

a) The impact of the information associated with the data sites. 

First, for the standard ConvTempNet and the plain encoder connected with a recurrent neural network, we observe that the classification performance of the same model on the data from a single site may noticeably vary depending on the site. This can be explained by the differences in data quality, degree of noise removal, and characteristics of the study participants, such as age and comorbidities. Furthermore, in both cases, the model trained on the joint multi-site dataset underperforms two out of four single-site models, which also indicates the negative effect of site-related variability on the model generalization ability. Even if several data sources were presented in a training sample, the model has difficulties in revealing robust and site-independent biomarkers of autism. This is additionally supported by the fact that model performance on the leave-one-site-out validation turns out to be extremely low and close to the random guessing. Thus, almost all the distinctive patterns learned by the model partly rely on the properties of a particular data site and are of no value in the analysis of data from a new source.

b) The effect of removing site-related information.

In the case of the models with fader network encoders connected with recurrent neural networks we observe that the performance on individual sites is in general slightly lower, especially for the model with strong elimination of site-related information. It is potentially caused by a slight distortion of latent features during adversarial training. However, on the multi-site data, these models demonstrate considerably better performance. Most significantly, the use of site-independent latent vectors improves the results in leave-one-site-out validation. ASD recognition on the completely new data from the previously unseen site is still less accurate than on the new data from sites presented in the training set. However, now the model noticeably outperforms the random guessing, which indicates that the disease biomarkers learned by it are more general and probably rely on the properties of the brain itself rather than on the properties of the data source.

It's also worth noting that a more thorough removal of site-specific variability does not improve the quality of the final model. Apparently, part of this variability turns out to be difficult to separate from the patient-specific patterns, and the attempt to completely eliminate it leads to the corruption of other information in latent vectors and the decrease in quality of the disease detection.



\begin{figure} [ht]
\begin{center}
\includegraphics[height=3.6cm]{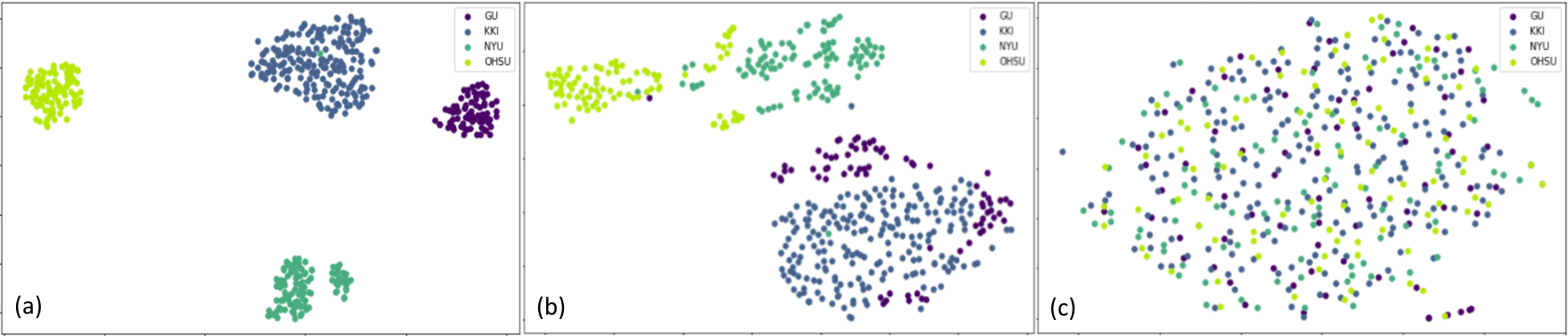}
\end{center}
\caption[example] 
{ \label{fig:tsne} 
 tSNE embedding of fMRI latent representations learned by (a) conventional autoencoder, (b) fader network with slight site-related information elimination, (c) fader network with strong site-related information elimination}
\end{figure} 

\subsection{Comparison of latent representations learned by conventional Autoencoder and Fader Networks}

Since the models with fader network encoders proved to be the most successful in the discussed above problem of autism recognition from multi-site full-size fMRI, here we visually inspect the latent representations that they extract from the brain images. We also compare them with the representations learned by the standard autoencoder in order to figure out how introducing the site-discriminator in the encoder training affects the features learned. 

For each subject in the considered dataset of samples from 4 largest sites in ABIDE-II, we encode it's 3D brain image for a single time-step, randomly sampled from fMRI sequence, into a latent vector using 1) the encoder of a plain autoencoder, 2) the encoder of fader network with slight elimination of site-related information, 3) the encoder of fader network with strong elimination of site-related information. After that, we reduce the dimensionality and plot the tSNE 2d-embedding for samples from all sites.

The results, presented in Figure \ref{fig:tsne}.a, indicate that for the plain autoencoder, latent representations contain information which allows clearly separating samples from different sites into distinct clusters. Figure \ref{fig:tsne}.b demonstrates that after we force the encoder to avoid preserving site - related knowledge by introducing discriminator into a latent space, the embeddings learned are tending to mix across the sites. In Figure \ref{fig:tsne}.c we observe a complete mixing of objects from different sites, which, however, is achieved at the cost of a partial loss of valuable information.

\section{Discussion}


We performed the first to our knowledge attempt toward the generalizable fMRI deep classification models. As we can see from the obtained results, the model performance can be significantly affected by the presence site-related variability in training data. However, we prove the ability of fader networks to be applied for construction site-invariant latent representations, which allow obtaining notable improvement of autism recognition, especially on the data from the previously unseen site. At the same time, the complete removal of any source-related part of the information is still undesirable, since it inevitably affects other patterns in the data that are important for pathology recognition.



\section{Acknowledgements} 
  The reported study was funded by RFBR according to the research project \textbf{20-37-90149}.



 

\bibliography{report} 

\begin{thebibliography}{10}

\bibitem{postema2019altered}
Postema, M.~C., Van~Rooij, D., Anagnostou, E., Arango, C., Auzias, G.,
  Behrmann, M., Busatto~Filho, G., Calderoni, S., Calvo, R., Daly, E., et~al.,
  ``Altered structural brain asymmetry in autism spectrum disorder in a study
  of 54 datasets,'' {\em Nature communications}~{\bf 10}(1),  1--12 (2019).

\bibitem{yahata2016small}
Yahata, N., Morimoto, J., Hashimoto, R., Lisi, G., Shibata, K., Kawakubo, Y.,
  Kuwabara, H., Kuroda, M., Yamada, T., Megumi, F., et~al., ``A small number of
  abnormal brain connections predicts adult autism spectrum disorder,'' {\em
  Nature communications}~{\bf 7}(1),  1--12 (2016).

\bibitem{abraham2017deriving}
Abraham, A., Milham, M.~P., Di~Martino, A., Craddock, R.~C., Samaras, D.,
  Thirion, B., and Varoquaux, G., ``Deriving reproducible biomarkers from
  multi-site resting-state data: An autism-based example,'' {\em
  NeuroImage}~{\bf 147},  736--745 (2017).

\bibitem{el2019hybrid}
El-Gazzar, A., Quaak, M., Cerliani, L., Bloem, P., van Wingen, G., and Thomas,
  R.~M., ``A hybrid 3dcnn and 3dc-lstm based model for 4d spatio-temporal fmri
  data: An abide autism classification study,'' in [{\em OR 2.0 Context-Aware
  Operating Theaters and Machine Learning in Clinical
  Neuroimaging}{\nolinebreak\hspace{0.1em}]},   95--102, Springer (2019).

\bibitem{kazeminejad2019topological}
Kazeminejad, A. and Sotero, R.~C., ``Topological properties of resting-state
  fmri functional networks improve machine learning-based autism
  classification,'' {\em Frontiers in neuroscience}~{\bf 12},  1018 (2019).

\bibitem{price2014multiple}
Price, T., Wee, C.-Y., Gao, W., and Shen, D., ``Multiple-network classification
  of childhood autism using functional connectivity dynamics,'' in [{\em
  International Conference on Medical Image Computing and Computer-Assisted
  Intervention}{\nolinebreak\hspace{0.1em}]},   177--184, Springer (2014).

\bibitem{cheng2017classification}
Cheng, D. and Liu, M., ``Classification of alzheimer’s disease by cascaded
  convolutional neural networks using pet images,'' in [{\em International
  Workshop on Machine Learning in Medical
  Imaging}{\nolinebreak\hspace{0.1em}]},   106--113, Springer (2017).

\bibitem{khosla2019ensemble}
Khosla, M., Jamison, K., Kuceyeski, A., and Sabuncu, M.~R., ``Ensemble learning
  with 3d convolutional neural networks for functional connectome-based
  prediction,'' {\em Neuroimage}~{\bf 199},  651--662 (2019).

\bibitem{li2020multi}
Li, X., Gu, Y., Dvornek, N., Staib, L., Ventola, P., and Duncan, J.~S.,
  ``Multi-site fmri analysis using privacy-preserving federated learning and
  domain adaptation: Abide results,'' {\em arXiv preprint arXiv:2001.05647}
  (2020).

\bibitem{dakka2017learning}
Dakka, J., Bashivan, P., Gheiratmand, M., Rish, I., Jha, S., and Greiner, R.,
  ``Learning neural markers of schizophrenia disorder using recurrent neural
  networks,'' {\em arXiv preprint arXiv:1712.00512}  (2017).

\bibitem{pominova2018voxelwise}
Pominova, M., Artemov, A., Sharaev, M., Kondrateva, E., Bernstein, A., and
  Burnaev, E., ``Voxelwise 3d convolutional and recurrent neural networks for
  epilepsy and depression diagnostics from structural and functional mri
  data,'' in [{\em 2018 IEEE International Conference on Data Mining Workshops
  (ICDMW)}{\nolinebreak\hspace{0.1em}]},   299--307, IEEE (2018).

\bibitem{di2017enhancing}
Di~Martino, A., O’connor, D., Chen, B., Alaerts, K., Anderson, J.~S., Assaf,
  M., Balsters, J.~H., Baxter, L., Beggiato, A., Bernaerts, S., et~al.,
  ``Enhancing studies of the connectome in autism using the autism brain
  imaging data exchange ii,'' {\em Scientific data}~{\bf 4}(1),  1--15 (2017).

\bibitem{craddock2013neuro}
Craddock, C., Benhajali, Y., Chu, C., Chouinard, F., Evans, A., Jakab, A.,
  Khundrakpam, B.~S., Lewis, J.~D., Li, Q., Milham, M., et~al., ``The neuro
  bureau preprocessing initiative: open sharing of preprocessed neuroimaging
  data and derivatives,'' {\em Neuroinformatics}~{\bf 4} (2013).

\bibitem{ganin2015unsupervised}
Ganin, Y. and Lempitsky, V., ``Unsupervised domain adaptation by
  backpropagation,'' in [{\em International conference on machine
  learning}{\nolinebreak\hspace{0.1em}]},   1180--1189, PMLR (2015).

\bibitem{lample2017fader}
Lample, G., Zeghidour, N., Usunier, N., Bordes, A., Denoyer, L., and Ranzato,
  M., ``Fader networks: Manipulating images by sliding attributes,'' in [{\em
  Advances in neural information processing
  systems}{\nolinebreak\hspace{0.1em}]},   5967--5976 (2017).

\end{thebibliography}
\bibliographystyle{spiebib} 

\end{document}